# Transient adhesion and conductance phenomena in initial nanoscale mechanical contacts between dissimilar metals

**William Paul, David Oliver, Yoichi Miyahara, and Peter Grütter**

Department of Physics, Faculty of Science, McGill University, Montreal, Canada.

E-mail: paulw@physics.mcgill.ca

**Abstract**
We report on transient adhesion and conductance phenomena associated with tip wetting in mechanical contacts produced by the indentation of a clean W(111) tip into a Au(111) surface. A combination of atomic force microscopy and scanning tunneling microscopy was used to carry out indentation and to image residual impressions in ultra-high vacuum. The ~7 nm radii tips used in these experiments were prepared and characterized by field ion microscopy in the same instrument. The very first indentations of the tungsten tips show larger conductance and pull-off adhesive forces than subsequent indentations. After ~30 indentations to a depth of ~1.7 nm, the maximum conductance and adhesion forces reach steady-state values approximately 12× and 6× smaller than their initial value. Indentation of W(111) tips into Cu(100) was also performed to investigate the universality of tip wetting phenomena with a different substrate. We propose a model from contact mechanics considerations which quantitatively reproduces the observed decay rate of the conductance and adhesion drops with a 1/e decay constant of 9-14 indentation cycles. The results show that the surface composition of an indenting tip plays an important role in defining the mechanical and electrical properties of indentation contacts.

## 1. Introduction

Nanoindentation is a widely used tool for the mechanical characterization of materials at small length scales [1–3]. However, as length scales decrease, the continuum description of contact mechanics breaks down due to several factors such as the roughness of indenters due to their crystalline nature [4,5], chemical bonding between materials causing adhesion [6], and the discretization of plastic processes [7,8]. Computational power permits the atomistic modeling of nanoindentation by molecular dynamics simulations up to some millions of atoms. These simulations are typically much smaller than experimentally accessible length scales (a million atoms are stored in a cube of side length ~25 nm), and the disparity in size hinders the comparison of observed phenomena between modeling and experiments.

Here, we combine field ion microscopy (FIM) with atomic force microscopy (AFM) and scanning tunneling microscopy (STM) to carry out atomic-scale nanoindentation using well-defined probes having dimensions matching those used in atomistic simulations. Well-controlled indentation at this length scale is made possible by this unique combination of techniques which allows for the atomic-scale characterization of the indentation probe, well-defined surface chemistry, force and conductance detection throughout indentation, and high resolution surface imaging of the resulting damage.

In this paper, we discuss recent experiments which indicate that the surface composition of indenters has a significant effect on the measured mechanical and electronic behaviour during indentation. These effects are only characterizable by indentation with a tip of known surface chemistry, which is accessible to us by preparation of a tungsten tip in FIM. The results imply that large quantitative discrepancies may exist between experiments and modeling if surface composition cannot be well controlled.

FIM is an indispensable preparation and characterization tool for nanoscale indenters because the atomic geometry of their apices can be imaged in real space, and the radius of the spherical tip can be interpreted from the image. Furthermore, the chemical composition of the apex is guaranteed: The removal of several layers of surface atoms by field evaporation in FIM is used to reveal a pristine tungsten surface. After that, the recently demonstrated 'force field' technique is used to ensure that impurities in the FIM imaging gas do not chemically react with the tungsten tip before it is used in the indentation experiment [9].

We find that the first mechanical contacts between a 'fresh' W(111) single crystal hemispherical tip and a Au(111) surface show unique behavior which stabilizes as material is transferred from the sample to the tip over repeated indentations. The very first indentations show much larger adhesion and maximal conductance, which we attribute to the change in the tip's surface chemistry and the build-up of gold on the tip. We report on the features of the force-displacement curves which might provide feedback to atomistic modeling of systems at the same length scale, and comment on challenges concerning the timescales of molecular dynamics simulations with regard to surface diffusion, which we believe plays a significant role in the observed behavior.

## 2. Experimental Methods

Experiments were carried out in ultra-high vacuum (UHV) at room temperature. Au(111) substrates were prepared by epitaxial growth of Au, 100 nm thick, on a ~1 mm x 5 mm piece of 50 μm thick mica. The Au(111) surface was cleaned by repeated 1 keV $Ne^+$ ion sputtering and annealing cycles in UHV to several cycles beyond the disappearance of carbon in Auger electron spectroscopy. The mica beam was cantilevered, such that it could deflect upon interaction with the tip. Forces were extracted with an absolute accuracy of ~10% by monitoring the cantilever's deflection by interferometry [10]. During indentation experiments, the conductance of the junction was recorded simultaneously with the load over the tunneling to contact range (pA to μA) using a logarithmic current preamplifier [11].

The indenter consists of a tip electrochemically etched from W(111) wire and prepared by flash annealing and degassing cycles in UHV [12,13]. Field evaporation was performed during FIM by raising the imaging field by ~10% relative to the field required for $He^+$ ion imaging of the apex. The clean W(111) tips prepared for these experiments had radii of 7.1 ± 0.7 nm and 7.2 ± 0.6 nm, determined by ring counting between pairs of (110), (111) and (211) planes [9,14–16]. The FIM images of these tip apices are shown inset in Figure 1(a) and (d). The apices of the W(111) tips end in three individually resolved atoms (trimer), similar to the tips described in Ref. [9].

**See end of manuscript for full-page figure**

**Figure 1:** Summary of the first 25 indentation curves of a fresh W(111) tip in a Au(111) substrate in two separate experiments. (a) & (d) Force upon loading. (b) & (e) Force upon unloading. (c) & (f) Conductance upon loading and unloading at -0.05 V sample bias. Insets show FIM images taken at (a) 6.3 kV and (b) 6.2 kV. Insets in (b) and (c) show time traces corresponding to the discontinuities observed between points A and B.

After preserving the atomic integrity of the tips using the 'force field' protocol while UHV conditions recovered after FIM [9], they were approached to tunneling interaction with Au(111) samples at a setpoint of 9 pA at –0.05 V sample bias, taking care that negligible overshoot of the tunneling setpoint occurred (thereby preventing tip crashes). Upon finding the sample surface, the tunneling setpoint was reduced to 3-4 pA at –0.25 V sample bias to minimize the frequency of spikes in the tunneling current (and thus transfer of material). This reduced setpoint was maintained under feedback for ~3 minutes in order for piezo creep to settle. The transfer of atoms from clean Au(111) samples happens spontaneously at room temperature, and is correlated to the occurrence of spikes in the tunneling current [9], therefore, before the first indent, the tip is expected to have some small amount of Au atoms at its apex. The tip was then moved under feedback to a new location on the sample for each indentation, where it was approached toward the sample by 3 nm from the tunneling condition which was 25 pA at –0.05 V (sample bias). The bias voltage was maintained during indentation in order to measure the junction conductance. The indentation sites were spaced by 25 nm in a 5 × 5 array. Arrays of indentations are very useful for the unambiguous identification of indentation sites due to the production of a regular pattern on the surface even in the presence of thermal drift, piezo hysteresis and creep which affect image linearity. The surface topography was not imaged before the indentations so that indentation could begin with a minimally modified tip.

## 3. Results

The resulting indentation curves from two such experiments are summarized in Figure 1 in the left and right columns. The general phenomena described here have been repeatedly observed in about a dozen experiments with different indenters – we present two sets of results with tips of nearly identical radii to emphasize the quantitative repeatability.

In Figure 1(a) and (d), we plot the force versus displacement during loading of the contact ('in' direction) for all 25 indentations, colour coded according to the legend inset in (a). The first loading curve is plotted in black over the coloured curves to emphasize the transient changes. Examination of the loading curves reveals that the rise of the repulsive force between the fresh tip (black curve) and the surface provides a left-hand bound to the subsequent indentations (subsequent curves show the onset of repulsive loading at a greater depth). The depth axis is zeroed to the distance of a 1 GΩ tunneling gap, so what we observe here is a displacement of the onset of repulsive contact *relative* to the point at which the tunneling current is measured. This is explained by the transfer and adhesion of gold atoms to the tip, adding a compressible yet conductive layer of variable thickness to the apex. The offset between the 1 GΩ tunneling gap and the onset of repulsive loading eventually becomes rather stochastic and reflects the compression of various amounts of substrate material (up to 3-4 Å) adhering to the tip before the rigid body W(111)/Au(111) contact occurs.

The corresponding unloading curves are shown in (b) and (e) for these experiments using the same color scheme. Upon unloading, the first curve in (b) shows an unloading stiffness of ~365 N/m. With successive indentations, the unloading contact stiffness decreases, and in the latter half of the curves it attains a more constant value of 292 ± 11 N/m (the uncertainty represents the standard deviation in the distribution of fitted stiffness values). The stiffness determination is not particularly accurate because the fit region must be manually selected to avoid events such as pop-ins / pop-outs, and adjustment of the fit region readily changes the determined values by ~5 %. The unloading curves for the first indentations with a nearly-pristine tip end with a massive attractive adhesion force reaching a maximal value of −157 nN in the first curve of (b). The adhesion rapidly reduces to a much smaller value of ~ −50 nN after several indentations.

The conductance of the junction measured during indentation is shown in (c) and (f) where both loading and unloading directions are plotted on the same graph (directions denoted by arrows). The loading conductance curves pass through the depth zero point at 1 GΩ during the tunneling regime. At maximum depth, the conductance generally reaches a near-maximum value. It is common for atomic rearrangements in the unloading curve to improve the junction conductance from its value at maximum depth by up to 20 %. For the first curve in (b) and (c), the junction conductance decreases abruptly from ~88 $G_0$ to ~18 $G_0$ at the point where the adhesion force jumps from −154 to −24 nN. This jump is labeled by points A and B in both the force (b) and current (c) plots. The sudden rearrangement is accompanied by a change in penetration depth due to the compliant cantilevered sample, so we plot a time trace of both force and current in the insets showing that this event is effectively instantaneous in time (here, limited by the bandwidth of our digital acquisition antialiasing filter of 512 Hz). The conductance decreases until the atomic wire breaks at a conductance of ~ 1 $G_0$ at a length of > 1.5 nm above the free surface depth.

The conductance of the first indentation reaches a maximum of 127 $G_0$. In the subsequent indentations, the conductance is markedly decreased and reaches a maximum of only 48 $G_0$ for the last indentation curve (see the progression of the curves near maximum depth in (c) and (f)). Also, hysteresis in the current channel in subsequent indentations is much smaller, showing that the atomic wire rearrangements happen earlier in the unloading process. These later curves often return to the tunneling regime after wire breakage, indicating that the contact has not separated to as large a distance as it did initially.

The surface damage was imaged with the same tip in STM after indentation, shown in Figure 2 (the skewed appearance of the indentation arrays is due to thermal drift and is enhanced by the fact that indentation was carried out from bottom to top, while imaging was carried out from top to bottom). One particular curiosity is the 18$^{th}$ indentation in experiment 2 which did not leave any permanent impression. A screw dislocation appears to occupy the site where the indentation was made – a pop-in was registered in the force-displacement curve, but apparently plasticity at this location was somehow reversed. Monoatomic islands of pile-up are located near most indentation sites, but in 8 of 50 locations, no pile-up is imaged. The pile-up could result from material ejected from the indentation site, from the wire neck breaking during unloading, or a combination of both effects.

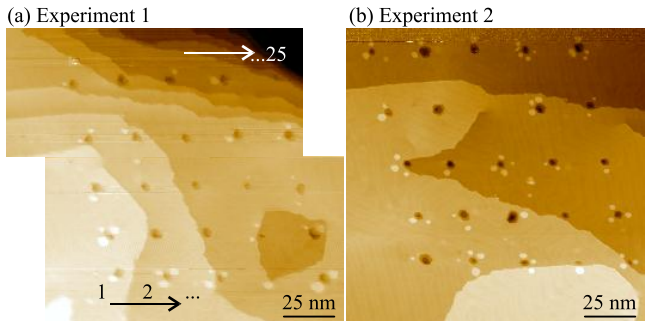

**Figure 2:** Topographic STM images of the indentation arrays produced in two experiments summarized in Figure 1. Both arrays were performed from left to right and bottom to top, as indicated by the arrows and numbers in (a). The top and bottom halves of image (a) are displaced due to a change in scan offset location carried out in the middle of the scan to allow the full array to be imaged. STM imaging conditions: (a) 12 pA, -0.05 V; (b) 10 pA, -1 V.

Hillocks have been observed at some distance from indentation sites (some tens of nm) in AFM-based indentation of KBr and Cu and have been interpreted as the intersection of plastic-induced dislocations with the surface [17,18]. STM imaging of the surface damage shows a lack of hillocks: around the indentation sites, the surface remains atomically flat with the exception of local rearrangements of the herringbone surface reconstruction. In our experiment, the indentation depth is comparatively shallow, and we estimate that the energy involved in creating the plastic damage in the substrate is of the order 100-200 eV (estimated by integrating the force-displacement curves for the most 'wetted' tips to avoid adhesion effects). This energy is rather low compared to the line energies of dislocations (~6 eV per atomic plane [19]). The energy budget and the lack of hillocks leads us to expect that the surface damage consists of a vacancy cluster after contact unloading.

The reduction of maximum conductance and adhesive force in these two experiments is plotted as a function of indentation number in Figure 3. The conductance and adhesion behavior is in good quantitative agreement between the two experiments with tips of similar radii, although the second experiment begins with an initially lower conductance and adhesion. It is difficult to say whether this deviation is statistically significant – defining an initial condition for the tips is hindered by the fact that their approach to tunneling proximity with the substrate allows for the transfer of material to the tip [9,20], though we have attempted to minimize this effect as described in the Experimental Methods section. Any discrepancies in the duration of tunneling or feedback conditions could result in a different initial condition of the tip apex, altered by material transfer.

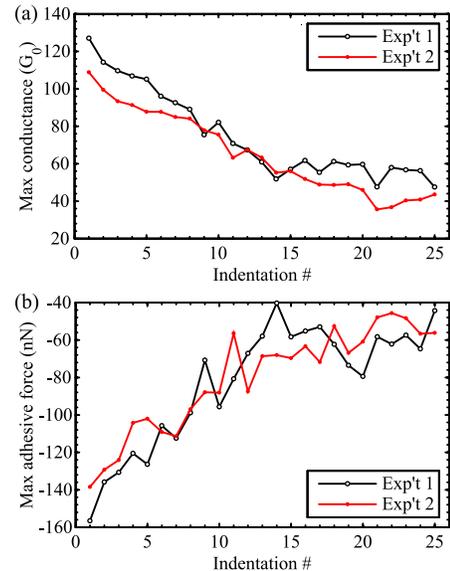

**Figure 3:** (a) Maximum conductance and (b) adhesive force measured in the two experiments shown in Figure 1.

One particularly intriguing feature about the transient behavior is that it takes a substantial number of mechanical contacts between the tip apex and gold surface to reach steady state. We hypothesize that it takes repeated indentations to reach equilibrium values conductance and adhesion because the transferred gold atoms rapidly diffuse away from the tip apex at room temperature [20]. After many indentations, however, a sufficient reservoir of gold may be established on the tip such that the behavior becomes repeatable.

The plateau in the conductance and adhesion behaviors beyond the first 25 indentations is illustrated for Experiment 1 in two subsequent 5 × 5 arrays carried out to 2.5 nm and 3 nm depth setpoints: Figure 4(a) and (b) show the maximum conductance and adhesive force, respectively. The conductance and adhesive force in the last 5 × 5 indentation array have average plateau values of $\langle G \rangle = 11 \pm 3\ G_0$ and $\langle F_{ahd} \rangle = -27 \pm 3$ nN, where the uncertainties express the standard deviation of the measured values. The steady state values represent a ~12× and ~6× drop from those of the first indentation.

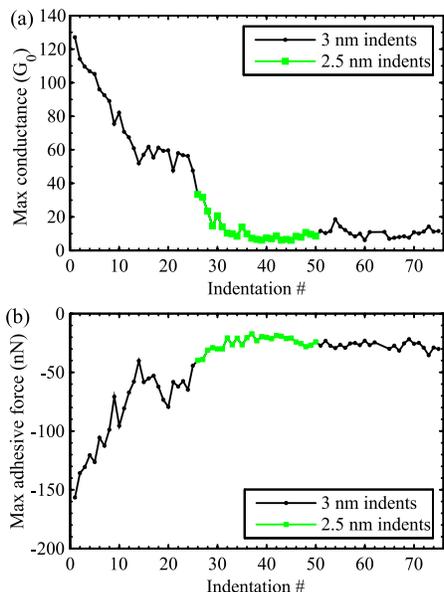

**Figure 4:** Maximum conductance (a) and adhesive force (b) tend toward a constant value after repeated indentation. Data shown is from Experiment 1, including two subsequent 5 × 5 indentation arrays to open-loop depths of 2.5 nm and 3 nm.

To investigate whether this phenomenon occurs in other metallic substrates, indentations were also performed on a Cu(100) single crystal with a W(111) tip of radius 7.2 ± 0.7 nm. Due to the fact that the sample was a 2 mm thick single crystal (i.e. not a compliant cantilever), forces could not be monitored during indentation – we report here only on the maximum junction conductance. In these indentations, a lower depth setpoint of 1.5 nm was chosen; this depth roughly corresponds to the maximum penetration depth in the Au(111) experiments (after subtracting for the compliant cantilever deflection).

A FIM image of the W(111) apex is shown inset in Figure 5 (a), and a STM topograph of typical surface damage resulting from indentation is shown in (b). The pileup is two Cu(100) atomic layers high, and the hole left in the substrate is two atomic layers deep. Both of these features reflect the 4-fold symmetry of the substrate (the outlines of these features are subject to tip convolutions in STM). As with the Au(111) surface, STM images show a lack of hillocks on the surface surrounding the indentation site; it is expected that the surface damage after unloading consists of a vacancy cluster.

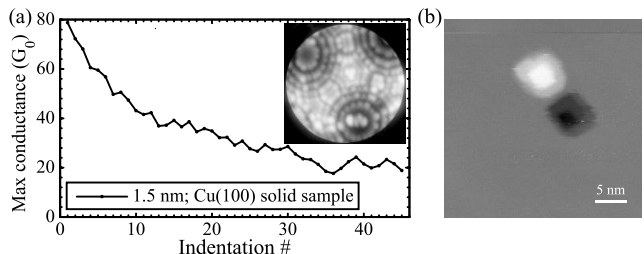

**Figure 5:** (a) Maximum conductance during indentation sequences on Cu(100). (b) STM image of the typical square shaped hole and pileup reflecting the symmetry of the substrate (9 pA, -0.8V). The vertical contrast in image (b) is 0.7 nm.

The maximum conductance of the first indentation reported in Figure 5(a) was ~80 $G_0$. After 25 indentations, the maximum conductance began to plateau around ~20 $G_0$, as shown in Figure 5(a). We note that a dozen STM images and 2 small indentations were performed before the indentations summarized in Figure 5(a), adding some uncertainty to the initial state of the tip. This result indicates that the 'fresh' tip would have had a conductance of at least 4× higher than the 'steady state' behavior obtained after tip wetting.

## 4. Discussion

In nanometer-scale indentations, such as those reported here, a sizable fraction of the atoms involved belong to the surfaces of the contacting materials. The high surface-to-volume ratio was advocated by Agraït et al. to be of importance at the 1-10 nm length scale as the work required to extend surfaces exceeds the work required to deform volumes [21]. Our studies indicate that the changing surface composition of the tip also affects adhesion, conductance, and contact stiffness during indentation. If the surface chemistry of the indenters is not well controlled, these important properties may not be accurately measured or modeled. The effects of surface composition observed here are not small: the adhesive force varies by ~6×, and conductance changes by more than an order of magnitude.

Below, we propose a model which reproduces the gradual saturation behavior, and we discuss possible physical mechanisms for the observed adhesion and conductance drops which open new directions for future experiments and modeling work.

### 4.1 Saturation behavior

The saturation of the maximum adhesion and conductance occurs after approximately the same number of indentations (including those performed on Cu(100)), suggesting that they reflect a common physical cause. We attribute the cause of these modifications from the initial condition of the tip to the wetting of the tip surface with substrate material. We reach this conclusion from the experimental evidence of wire-drawing and the variable onset of repulsive contact loading with respect to the tunneling point. Results of molecular dynamics simulations for the same materials system also support the idea that gold is transferred to the tungsten tip during the indentation process [22].

Why does it require roughly 30 contacts with the surface to reach saturation? We speculate that the tip gradually builds up a surface density of gold which eventually reaches a saturation point as indentations progress. This could occur if only a finite region near the apex of the tip is capable of holding the transferred material. The field evaporation cleaning of the tungsten tip will serve to create clean surfaces in the vicinity of the FIM-imaging region, but only where the electric field is highest close to the end of the tip, not all the way up the tip shank. The tip shank is expected to be completely chemisorbed with dissociated rest gases ($N_2$, $O_2$, CO) after several days in vacuum [9], and the added surface roughness in this region should inhibit the diffusion of transferred material away from the field-evaporation cleaned apex.

A simple model for such a wetting scenario is suggested: First, assume that the transferred gold atoms are confined to the hemispherical end of the tip which has been cleaned by field evaporation and cannot diffuse further up the shank due to its rough surface. During each contact between the tip and substrate, a full layer of gold wets the tip in the area which was in contact during indentation. Once the tip is retracted, thermally activated surface diffusion takes place fast enough (compared to the time delay between indentations) to rearrange the material homogeneously over the tip. The next indentation transfers another complete gold layer to the tip within the contact area. Upon retraction, the gold rearranges homogeneously over the tip. As indentations are repeated, the density of gold atoms tends toward a uniform coverage on the tip. We illustrate the transferred gold within the relevant contact area on a ball model of the tip in Figure 6. This contact area makes up a small fraction of the tip hemisphere which has been cleaned by field evaporation.

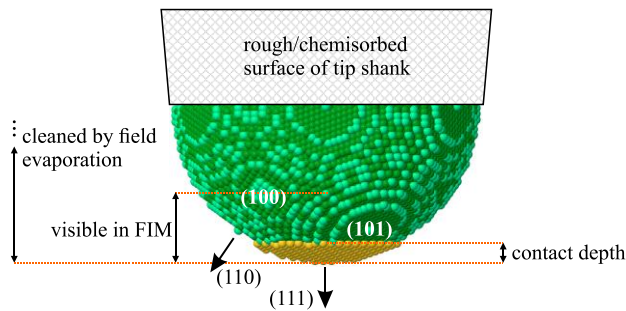

**Figure 6:** Side view of an atomic model of a 7.1 nm radius W(111) tip apex. Low-index planes are identified. The apex of the tip contains added gold atoms up to the depth of indentations in these experiments. The approximate region of the tip visible in our FIM images is identified (our field of view is limited by an aperture in the microscope housing). Field evaporation cleaning of the tip guarantees a clean tungsten surface up to about the start of the tip shank.

Our observations of transferred atoms from Au(111) surfaces to W(111) FIM tips at 298 K and 158 K show that activation energies for diffusion over the tip surface are low enough to permit complete rearrangement of the atoms over all visible regions of the tip (in FIM) at room temperature, but the diffusion is frozen out at 158 K [20]. This justifies our assumption that the transferred atoms can redistribute themselves by surface diffusion between indentations at room temperature.

If the first contact transfers a layer of gold to a fraction of the available surface area $a$, after diffusion has rearranged the gold layer uniformly, the density of gold on the tip after the first indentation will be $d_1 = a$. The density after the second indentation will be $d_2 = a + (1-a)d_1$. After $n$ indentations, we have a gold atom density of

$$d_n = a + (1-a)d_{n-1}. \qquad (1)$$

This recursion relation can be rewritten as a difference equation and integrated to yield an exponentially saturating density of

$$d = 1 - \exp(-an), \qquad (2)$$

where $n$ is the indentation cycle number. Eq. (2) shows that the decay constant of the saturation is $a$, corresponding to the surface area fraction of the tip apex picking up gold during indentation. We note that the continuous description of the recursion relation (Eq. (1)) is only valid for small $a$, but the error is less than 10% for $a < 0.18$.

The contact area fraction is obtained by dividing the contact surface area during indentation by the total area available for the gold to diffuse (assumed here to be a half-sphere – the tip shank has a shallow cone half-angle of only ~5°):

$$a = \frac{A_{contact}}{2\pi R_{tip}^2}. \qquad (3)$$

The contact area during indentation can be estimated using contact mechanics models [23]. With an elasticity parameter of $\lambda \approx 0.4$ for the tungsten-gold contact, the adhesive behavior should be in between JKR and DMT models [24,25]. These models should represent upper and lower bounds for the contact area given the maximal forces measured during indentation (for our estimations, a combined modulus of 80 GPa is used for the W-Au contact [26], the surface energy is taken to be 1 J/m$^2$ in the JKR model, the tip radius determined by FIM is 7.1 nm, and the maximal force is ~225 nN).

From these models, we estimate a surface area fraction between 0.07-0.11 which suggests a decay constant ($1/a$) of 9-14 indentation cycles. This is in very good agreement with the observed conductance and adhesion behavior which in the 5 experiments reported (including Cu(100)) have decay constants in the range of 11-14 cycles obtained by exponential fits.

Although some significant assumptions went into our estimation (such as the half-sphere available area for material diffusion and that adatom coverage has the same atom density and thickness over the tip), we obtain a reasonable quantitative understanding of how many indentation cycles are necessary to achieve a wetted tip. To further test such a model, future work could investigate low temperature tip-wetting where the adatom diffusion over the tip surface is frozen out. Additionally, the deliberate alteration of the contact surface area fraction $a$ could be achieved by using tips of differing radii and investigating several indentation depths.

*4.2 Adhesion drop*

We believe that the added gold on the tip in the latter indentations serves to lower adhesion by facilitating atomic rearrangements upon retraction of the tip. Rather than the gold substrate adhering directly to the tungsten tip, the layer of gold mediates the rearrangement of atoms during pull-off. Additionally, the gold wetting layer alters the surface chemistry of the tungsten tip which in turn changes the relevant diffusion barriers.

Recently, Gosvami *et al.* observed that the onset of wear in friction force microscopy studies on gold surfaces occurs much more readily below 177 K than at room temperature [27]. The mechanism by which friction and wear were reduced at room temperature was attributed to the mobile adatom population serving to rearrange and renew the sliding contact. We believe that our observation of reduced adhesion stems from similar diffusion effects, which become more pronounced with the gradually saturating gold atom population on the tip.

We have attempted to investigate the effect of wetted tips on adhesion in molecular dynamics simulations by indenting a gold-covered tungsten tip into a fresh gold slab, however no significant change in the pull-off force was apparent. With the indentations advancing 0.25 Å in 2.5 ps, followed by a 5 ps rest time for equilibration [22], there is little time for diffusion to occur. The probability of overcoming diffusion barriers of some tenths of an eV are very small in the period of one lattice vibration, even at room temperature (the Boltzmann factor $\exp(-E_a/k_B T)$ gives ~ 1/50 and 1/2200 for activation energies of 0.1 and 0.2 eV, respectively). Our experiments, however, are carried out about $10^9$ times slower: with a total time of several seconds per indentation, these diffusion events can easily take place.

This is an important example of how molecular dynamics can fail to reproduce observed phenomena in experiments, such as adhesive forces, due to the disparity in timescales.

*4.3 Conductance drop*

We have previously investigated the conductance loss due to interfaces between dissimilar metals, showing that the conductance across a W(111)/Au(111) boundary is diminished by a factor of 4× due to back-reflection at the interface due to the poor overlap of Bloch states [23]. Ballistic transmission losses at interfaces and grain boundaries have been investigated in relatively few experiments [28]. Experimental control over contact area and grain orientation have left this subject more suited to theoretical investigations [28–31]. The common theme of these studies is that interfaces between different materials, grain boundaries, disorder, and vacancies contribute to additional electrical resistance due to either backscattering or local disruptions in potential.

As the tip becomes wetted with gold, we expect that the system builds additional interfaces with different orientations which will add to ballistic scattering. Although a first principles calculation of the equilibrium configuration of gold on the large faceted tip surface is unfeasible for a system of this size, we speculate on what some of the interface structures might be.

There is a severe geometric mismatch between the bcc tungsten lattice of the tip and the fcc gold lattice of the substrate. The close-packed plane of the bcc tungsten tip is the W(110) plane; three of which exist at a ~35° angle from the W(111) apex (the (110) and (101) planes are indicated in Figure 6). Much of the tip surface near the apex actually consists of stepped (110) surfaces. These close-packed (110) planes provide the best lattice match to Au(111), therefore one could expect them to support a (111)-like layer of gold after multiple indentations saturate the gold ad-layer density.

Thin films of Au have been studied on W(110) surfaces by angle-resolved ultraviolet photoelectron spectroscopy (ARUPS) and have shown bulk-like Au(111) behavior with the existence of Fabry-Pérot-like quantum-well states above 3 atomic layers [32]. Below 2 atomic layers, the gold layer consists of slightly disordered Au(111)-like layers with electronic structure which deviates strongly from a perfect Au(111) surface. We expect these concerns of epitaxy and lattice strain to affect electronic transport through the tip-sample junction once the tip has become wetted.

If the Au(111)-like surfaces supported by the three W(110), W(101) and W(011) planes surrounding the W(111) apex were to meet at the apex, they would not be able to join to form a lattice-commensurate gold crystal. This leads us to deduce that there will be some form of crystalline frustration near the W(111) apex. The Au(111)-like planes on the W(110) facets are also oriented differently than those of the Au(111) substrate. Building up the tip interface layer is therefore expected to add mismatched interfaces to the system which would act to hinder ballistic transmission.

A diagram of the relevant interfaces and disorder structures which are expected to affect ballistic transport during indentation is shown in Figure 7. The quasi-Au(111) on the W(110) plane is not in registry with the Au(111) substrate, nor with the Au(111)-like layers on the other W(110)-type planes (W(101) and W(011), not shown). A disordered region at the apex of the W(111) tip and Au(111) surface may also occur from lattice mismatch and indentation stresses. Other concerns such as vacancies and dislocations are also illustrated which have been addressed in previous work regarding the junction conductance of a loaded contact [23].

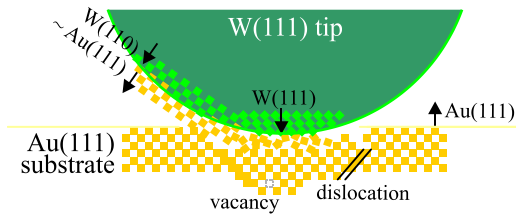

**Figure 7:** Illustration of incommensurate lattice matching between quasi-Au(111) on the W(110) planes, the W(111) apex and the Au(111) surface which may be enhanced upon tip wetting adding electron scattering sites to the junction. Dislocations, vacancies and disorder are also expected to affect junction conductance.

## 5. Conclusion

Tip wetting of fresh W(111) tips is observed during indentation contacts with Au(111) substrates. Evidence of tip wetting includes wire pulling, strong adhesion, and variably thick layers of conductive material transferred to the W(111) apex as (determined by the offset of repulsive indentation loading with respect to tunneling currents). The eventual saturation of the density of gold atoms on the tungsten tip surface leads to the transient behavior observed in the maximum adhesion and conductance of the junction. Though we cannot exclude the possibility of contaminant atoms affecting the adhesion and conductance behaviors, we do not expect this effect to be dominant due to the respectable UHV conditions corroborated by the clean surfaces and step edges in STM.

The adhesion and conductance maxima fell off with a decay constant of 11-14 cycles in our experiments on Au(111) and Cu(100). A simple model which assumes full coverage of gold atoms within the contact area and their subsequent diffusion over the tip between indentations predicts a decay constant in the range of 9-14 cycles for our tips and indentation depths. For the Au(111) substrate, the conductance and adhesion values drop by about 12× and 6× respectively from their initial values. Conductance on the Cu(100) substrate drops by at least 4×.

The size scale of these indentations is in direct correspondence with what is possible in molecular dynamics simulations. However, we caution that surface diffusion processes are likely to play a significant role in defining phenomena such as adhesion in experiments at room temperature. The discrepancy in time-scales of simulations cannot reproduce these features seen in experiments.

The strong dependence of adhesion, conductance, and unloading stiffness on the wetted state of the tip surface highlights the importance of controlling the indenter's surface composition for quantitative investigation of mechanical properties by indentation. We expect that the preparation of indenters by FIM will be of great importance to understanding mechanical contacts on the atomic length scale.

## Acknowledgments

Funding from NSERC, CIFAR, and RQMP is gratefully acknowledged.

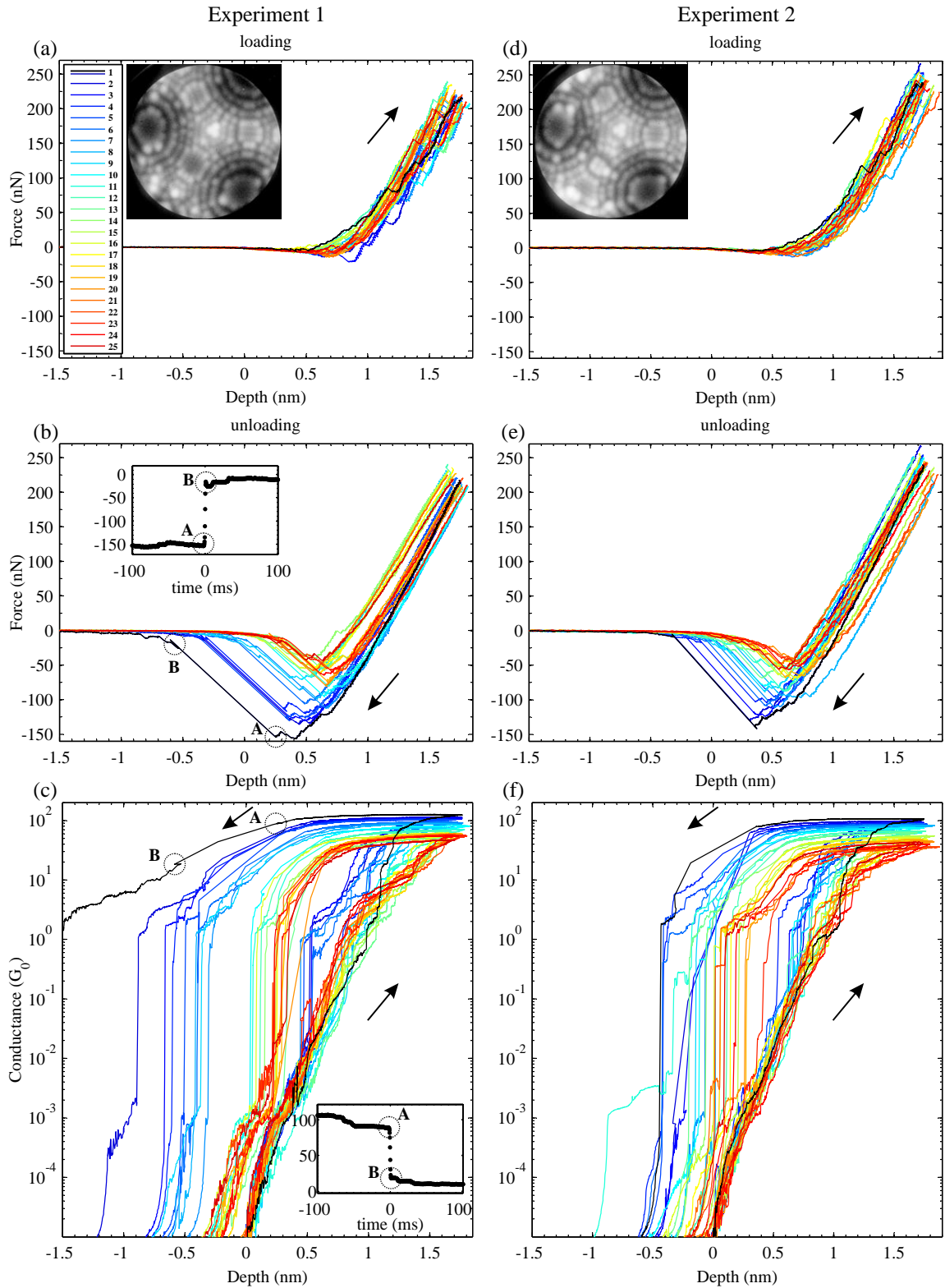

**Figure 1:** Summary of the first 25 indentation curves of a fresh W(111) tip in a Au(111) substrate in two separate experiments. (a) & (d) Force upon loading. (b) & (e) Force upon unloading. (c) & (f) Conductance upon loading and unloading at -0.05 V sample bias. Insets show FIM images taken at (a) 6.3 kV and (b) 6.2 kV. Insets in (b) and (c) show time traces corresponding to the discontinuities observed between points A and B.